\documentclass{kluwer}    
\input{epsf.tex}

\newdisplay{guess}{Conjecture}

\begin{document}                                                              
                   
\begin{article}
\begin{opening}    
\title{Should elliptical galaxies be idealised as collisionless equilibria?}
\author{Henry E. \surname{Kandrup}\email{kandrup@astro.ufl.edu}}
\institute{Department of Astronomy, Department of Physics, and Institute
for Fundamental Theory, University of Florida, Gainesville, 
Florida, USA 32611}
\runningauthor{H. E. Kandrup}
\runningtitle{Should elliptical galaxies be idealised as collisionless 
equilibria?}

\date{\today}

\begin{abstract}
This review summarises several different lines of argument suggesting that
one should not expect cuspy nonaxisymmetric galaxies to exist as robust, 
long-lived collisionless equilibria, {\it i.e.,} that such objects should not 
be idealised as time-independent solutions to the 
collisionless Boltzmann equation. 
\end{abstract}

\end{opening}

\section{Motivation}
Should elliptical galaxies be visualised as objects that spend most of their
lives in or near a true equilibrium, continually disturbed somewhat by nearby
objects, but not exhibiting any systematic dynamical evolution? Or should they
be viewed instead as objects that may be close to some equilibrium, but are
drifting through phase space in such a fashion as to manifest a systematic
secular evolution on time scales shorter than $t_{H}$, the age of the Universe?
In other words: once they have settled down towards a near-equilibrium, should
galaxies be viewed as static or dynamic entities?

Over the age of the Universe, ellipticals clearly change in terms of such
properties as colour. However, it is probably fair to say that, until recently,
many, if not most, astronomers have typically ignored the possibility of
systematic dynamical changes except, perhaps, in response to very close
galaxy-galaxy encounters, arguing, {\em e.g.,} that discreteness effects
({\em i.e.,} gravitational Rutherford scattering between individual stars) 
should be unimportant since the natural relaxation time 
$t_{R}{\;}{\gg}{\;}t_{H}$.
The object of this review is to argue that this conventional wisdom may not be
completely correct. Complex equilibria, especially cuspy and/or nonaxisymmetric
equilibria, may not exist or may be very hard for a real galaxy to find; and,
especially in high density environments, external irregularities may be 
significantly more important in triggering systematic changes in the bulk 
structure of a galaxy than has been generally recognised  hitherto.

\section{Can realistic equilibrium models exist?}
Dating back to the pioneering work of Eddington and Jeans, galactic dynamicists
have typically tried to construct equilibrium solutions to the collisionless
Boltzmann equation ({\em CBE}) using global integrals 
like the energy $E$ and rotational angular momentum $J_{z}$. Since these
quantities are conserved, any function of them will also be conserved, so that
they can be used to define an equilibrium phase space density, 
or distribution function, $f_{0}$. However, such 
`standard' equilibria are problematic if one allows for the possibility
that many galaxies are at least moderately triaxial, as one seems compelled to
do from an analysis of observations ({\em cf.} \citeauthor{TaM} \citeyear{TaM})
Generic time-independent, albeit nonaxisymmetric, potentials admit only one 
continuous symmetry, namely time-translation, so that there is only one global 
integral. If the system is nonrotating, this corresponds to the energy $E$; 
if, instead, the system is in uniform rotation, this corresponds to the Jacobi 
integral $E_{J}$.

The problem, then, is that it seems impossible to model triaxial configurations
with a strong central condensation in terms of equilibrium solutions 
$f_{0}(E)$ or $f_{0}(E_{J})$. One knows, {\em e.g.,} that every $f_{0}(E)$ 
must correspond to a spherically symmetric configuration 
(\citeauthor{PaA} \citeyear{PaA}).
A rotating solution $f_{0}(E_{J})$ {\em can} in principle correspond to a 
triaxial configuration, but such configurations seem unrealistic ({\em cf.} 
\citeauthor{pov1} \citeyear{pov1}, \citeauthor{IaM} \citeyear{IaM}).
Physically these triaxial 
configurations can exist for the same reason as the triaxial Riemann 
ellipsoids, namely because it is energetically favourable for the 
configuration to deviate from axisymmetry. However, this requires rotation 
rates that are too large to be realistic and central condensations much 
smaller than those observed in real early-type galaxies ({\em cf.} 
\citeauthor{lau} \citeyear{lau}).
For example, triaxial equilibria with distribution
functions appropriate for a rotating polytrope cannot have a central density 
larger than about $3.12$ times the mean density.

One can of course seek to construct models with `unobvious' symmetries such 
as, {\em e.g.,} models corresponding to integrable St\"ackel potentials. 
However, nonaxisymmetric models admitting two or three global integrals are 
extremely nongeneric, possessing continuous 
symmetries whose physical origins are far from obvious. Three-integral 
potentials constitute a set of measure zero in the set of genuinely 
three-dimensional potentials and, as such, if one wants to model real galaxies 
with such integrable potentials, it would seem crucial to identify a guiding 
principle which would explain why it is that nature selects such equilibria. 
One might perhaps hope to model triaxial galaxies in terms of potentials that 
admit only two global integrals. However, it is not clear whether there exist 
realistic potentials that admit only two global integrals which are not 
axisymmetric; and, even presuming that such potentials exist, these would 
again seem comparatively nongeneric.

Alternatively, one can look for equilibria involving `local integrals' ({\em
cf.} \citeauthor{LaL2} \citeyear{LaL2}),
the conserved quantities that make 
regular orbits behave as if they were integrable, even if the potential is 
nonintegrable and chaotic orbits also exist 
(\citeauthor{kan98a} \citeyear{kan98a}).
One can, 
{\em e.g.,} try to construct equilibria which contain both regular and chaotic
orbits but which, for fixed $E$ or $E_{J}$, assign different weights to 
the regular and chaotic phase space regions. This is, in fact, the tack 
implicit in almost 
all work with nonintegrable potentials that break axisymmetry, including the 
triaxial generalisations of the Dehnen (\citeyear{deh}) potentials considered, 
{\em e.g.,} by Merritt and Fridman (\citeyear{MaF}).

This problem can be addressed numerically using some variant of 
Schwarzschild's (\citeyear{sch}) 
method. What this entails is (i) specifying the presumed mass density 
${\rho}$ and gravitational potential ${\Phi}$ for some time-independent 
equilibrium,
(ii) generating a huge library of orbits evolved in this potential, and
then (iii) trying to select a weighted ensemble of orbits from that library
which reproduces the assumed density. Only by demanding that this weighted
ensemble reproduce the original ${\rho}$ can one ensure that one has a true
self-consistent equilibrium, in which the matter in the galaxy evolves in the
potential generated by the matter itself.

The usual implementation of this method might seem problematic in that it
ignores the role of conserved quantities like energy. However, it was shown 
some years ago (\citeauthor{pov2} \citeyear{pov2})
that these quantities are really hiding in
the method, at least for the special case of integrable potentials, and
more recently Kandrup (\citeyear{kan98a}) 
has shown that this method can also be modified
in a natural fashion to incorporate `local integrals'. The key observation
is that it is not orbits
{\em per se} that should be considered as the fundamental ingredients. Rather,
the crucial point is to select a collection of time-independent building blocks
which, being time-independent, can be used as static constituents for a 
time-independent equilibrium. Very recently, \citeauthor{CZZ} (\citeyear{CZZ})
have 
shown how, for the special case of an integrable potential, one can actually 
proceed semi-analytically, provided only that action-angle variables can be 
implemented explicitly: There is a one-to-one correspondence between values of 
the actions and time-independent density building blocks; and, by sweeping 
through all possible values of the actions, one is guaranteed to consider all 
possible time-independent building blocks. However, it does not seem
possible to generalise this approach to nonintegrable potentials. 

Unfortunately, there are obvious problems with Schwarzschild's met-hod or any
other numerical scheme. 
In particular, there is no proof that, for any given
${\rho}$, a self-consistent equilibrium exists or that that equilibrium
is unique. Indeed, even assuming that exact solutions do exist, there is no
guarantee that a numerical `solution' is a reasonable approximation to some
real solution. Whether or not an equilibrium exists, the numerical prescription
will find some `best fit' solution, and there is no reason {\em a priori} to
assume that that `best fit' corresponds to a {\em bona fide} collisionless
equilibrium. Alternatively, the inability to generate anything remotely 
resembling a `reasonable' solution need not guarantee that no solution exists:
this may simply signal an incomplete orbital library. At the present time, the
best that one might hope to do is sample one's purported equilibria to generate
$N$-body realisations, evolve these $N$-body realisations into the future, and
then determine whether these behave more or less stably for a finite time. 
However, even this is problematic. The normal discretisation involved in
implementing Schwarzschild's method is so coarse that $N$-body realisations
generated from Schwarzschild models of stable equilibria like Plummer spheres
can behave unstably \citeauthor{siop} (\citeyear{siop}). 

Another, potentially even more serious, problem with almost all work hitherto
is that the density distributions which have been considered are highly
idealised.
Most Schwarzschild modeling has involved an assumption of strict axisymmetry
or strict ellipsoidal symmetry with constant axis ratios. For example, the
claim that triaxial equilibria cannot exist for galaxies with very steep cusps 
is based almost completely on an analysis of the triaxial generalisations of 
the Dehnen potentials 
(\citeauthor{MaF} \citeyear{MaF}). However, it is by now well 
established that real ellipticals tend to have distinctly disky or boxy 
isophotes, the details of which correlate with other properties of the galaxy,
such as the steepness of the central density cusp or the bulk rotation rate
({\em cf.} \citeauthor{KaB} \citeyear{KaB}).
Moreover, even if a galaxy is nearly 
axisymmetric in the center, as seems likely for the coreless ellipticals,
they could be distinctly non-axisymmetric in their outer regions. The claim 
that triaxial potentials containing a very large supermassive black hole have 
`too many' chaotic orbits ({\em cf.} \citeauthor{MaV2} \citeyear{MaV2})
may well reflect
unnatural attempts to combine potentials with incompatible symmetries 
({\em cf.} \citeauthor{KaSd} \citeyear{KaSd}):
the black hole is presumably nearly
spherical or axisymmetric, but one typically assumes that the surrounding 
galaxy is triaxial with fixed axis ratios down to very small scales.

In any event, the fact that real galaxies are disky or boxy is probably no
accident, and it would seem that realistic galactic models should become more
nearly axisymmetric towards the center. However, such `more complex' objects 
might seem even less likely to manifest continuous symmetries that give rise
to global integrals, so that, assuming that they exist, equilibria with such
shapes would be even more likely to rely on `local integrals'.
\section{Will real galaxies evolve towards such equilibria?}
Over the past several decades, considerable effort, both analytic and
numerical, has been devoted to the construction of equilibrium 
solutions to the collisionless Boltzmann equation. Unfortunately, 
however, much less is known about a time-dependent evolution governed by the
{\em CBE}. There is, {\em e.g.,} no proof that generic initial data will 
evolve towards a time-independent equilibrium, even assuming that the 
configuration is gravitationally bound. Even such a basic property as {\em 
global existence}, {\em i.e.,} the fact that $f(t)$ does not evolve 
singularities and/or caustics, was only proven in the early 1990's 
(\citeauthor{pfaf} \citeyear{pfaf}, \citeauthor{scha} \citeyear{scha}).
It would seem that the only hard results 
about a time-dependent evolution that have been established to date concern 
the behaviour of quantities like time-averaged moments in an asymptotic 
$t\to\infty$ limit ({\em cf.} \citeauthor{batt} \citeyear{batt}).
That so little is known is not really
surprising. Even for the seemingly simple case of mechanical systems with
short range forces, it is often very difficult, if not impossible, to prove
that there is any approach towards equilibrium 
({\em cf.} \citeauthor{yas} \citeyear{yas}).

In this regard, it should be emphasised that there exist exact time-dependent
solutions to the {\em CBE} which do {\em not} manifest any approach towards an
equilibrium.  For example, \citeauthor{LaG} (\citeyear{LaG}) used semi-analytic
techniques to construct a solution which corresponds to finite amplitude, 
undamped oscillations about an otherwise time-independent equilibrium $f_{0}$.
Moreover, at least for the toy model of one-dimensional gravity, 
counter-streaming initial conditions, corresponding to a collision
between two galaxies initially in equilibrium, can yield a numerical evolution
towards a final state with undamped oscillations ({\em cf.} 
\citeauthor{MFR} \citeyear{MFR}).
{\em A priori} it might seem surprising that such oscillations do 
not exhibit linear and/or nonlinear Landau damping which would cause them to 
phase mix away. The crucial point physically is that these solutions contain 
`phase space holes,' {\em i.e.,} regions in the middle of the otherwise 
occupied phase space regions where $f\to 0$, so that one has the possibility 
of excitations that do not undergo a particle-wave interaction. 

The idea that oscillations never damp may seem too extreme to be realistic. 
However these results would suggest that internal irregularities could persist 
much longer than the theorist is wont to assume.

In any event, because an evolution governed by the {\em CBE} is Hamiltonian
({\em cf.} \citeauthor{kan98c} \citeyear{kan98c}), 
any statement regarding an approach towards
equilibrium must entail a coarse-grained description of the system. This could,
{\em e.g.,} involve a consideration of coarse-grained distribution functions, 
as in the original proposal of violent relaxation (Lynden-Bell 1967). 
Alternatively, one might 
consider the evolution of a collection of
lower order moments, {\em e.g.,} in the context of a cumulant expansion, an 
approach well known from plasma physics and accelerator dynamics. In either 
setting, the obvious question is: how fast, and how completely, do the 
observables of interest -- either coarse-grained distributions or lower order 
moments -- approach time-independent values? 

In a galaxy that is far from equilibrium, it is not unlikely that many orbits 
will exhibit an exponentially sensitive dependence on initial conditions, and 
the resulting {\em chaotic mixing} ({\em cf.} \citeauthor{mah} \citeyear{mah},
\citeauthor{MaV1} \citeyear{MaV1}, \citeauthor{kan98b} \citeyear{kan98b})
will certainly help a galaxy shuffle itself up. However, 
there is no guarantee that the galaxy will settle down all that efficiently 
towards a time-independent or nearly time-independent equilibrium! Indeed,
recent numerical simulations ({\em cf.} \citeauthor{VaW} \citeyear{VaW})
suggest 
that close encounters between galaxies can lead to long-lived pulsations.
(See also the 
oscillations described in \citeauthor{Dick} \citeyear{Dick}.)
It will be argued below that such oscillations could in fact trigger secular 
evolution on a time scale $<t_{H}$.

Analyses of flows in a fixed potential suggest 
({\em cf.} \citeauthor{kan98b} \citeyear{kan98b}) that,
when evolved into the future, generic ensembles of initial conditions {\em do} 
eventually exhibit a coarse-grained approach towards equilibrium. However, 
this can be an extremely complex, multi-stage process for nonintegrable 
potentials with a phase space that admits a complex coexistence of both 
regular and chaotic regions and, consequently, is riddled by a complex Arnold 
web. In particular, it is apparent that the rate of approach 
towards equilibrium can depend sensitively on the level of coarse-graining: 
probing the system at different scales and/or in terms of different order 
moments can lead to significant differences in the rate associated with any 
approach towards equilibrium. 
This is hardly surprising given 
the physical expectation that, as a result of phase mixing, power should 
cascade from larger to shorter scales.

The obvious point in all this is that great care must be taken in deciding what
one ought to mean by asserting that a galaxy is `nearly in equilibrium.' A
galaxy could, {\em e.g.,} look `nearly in equilibrium' when viewed in terms of
its bulk properties, but still exhibit significant shorter scale variability
and, most importantly, still be distinctly `out of equilibrium' from the
standpoint of dynamics. 

A crucial point, then, is that systems that have not achieved a true
time-independent equilibrium tend to be more susceptible to external stimuli
than systems that are in a true equilibrium. This is especially true for
complex systems characterised by a six-dimensional phase space that admits 
a complex coexistence of regular and chaotic regions, with structures like
bars or cusps that rely on an intricate balance of `sticky'
({\em cf.} \citeauthor{con} \citeyear{con}) and wildly chaotic
orbits.

Real galaxies, of course, are not strictly collisionless, and it is clear that
dissipative gas dynamics must have played an important role in the earliest
stages of galaxy formation where, even allowing for large quantities of 
nonbaryonic dark matter, much of the (proto-)galaxy is comprised of dissipative 
gas which has not yet been converted into stars. The obvious point, then, is 
that dissipative effects associated with this gas could play an important role 
in driving the system towards a true equilibrium. Indeed, such dissipative 
effects, which should be more important in the higher density central regions, 
might drive the central regions of a cuspy galaxy towards a state that is much 
more nearly axisymmetric than the lower density, outer portions of the galaxy.

Nevertheless, even if dissipation is a crucial element for the evolution
of primordial equilibria or near-equilibria, it would seem comparatively
unimportant when considering the effects of recent collisions and other close 
encounters between ellipticals. Over the past decade or so, it has been 
recognised that ellipticals are not as gas-poor as was originally believed. 
However, most of the gas in ellipticals exists at high temperatures and low 
densities, so that its dissipative effects should be minimal. It seems 
implausible that dissipative gas dynamics could play a dominant role in how 
elliptical galaxies readjust themselves after a strong encounter with another 
galaxy.
 
However, dissipative gas dynamics is not the only physical effect which is
ignored by the {\em CBE}. Real galaxies are also subjected to a variety of 
other perturbations which could in principle be important. For example, as
stressed, {\em e.g.,} by 
\citeauthor{MaF} (\citeyear{MaF}), the central regions of
cuspy galaxies are so dense that the relaxation time $t_{R}$ associated with
gravitational Rutherford scattering between neighbouring stars can be less
than or comparable to $t_{H}$. And similarly, galaxies are subjected 
continually to perturbations reflecting the effects of companion galaxies and 
other nearby objects which, especially in high density clusters, can be
appreciable.

An obvious question, therefore, is: what are the potential effects of ongoing
low amplitude perturbations? On the one hand, one might argue that they will
interfere with a systematic approach towards equilibrium since they imply that
the density distribution necessarily varies in time. On the other, one might
argue that these perturbations actually expedite the approach towards 
equilibrium, since the time-varying forces to which the galaxy is subjected
could facilitate violent relaxation by accelerating phase space transport.
In particular, it would not seem completely implausible to argue that, in
many cases, galactic evolution should be viewed as a two-stage process. Early 
on a galaxy could have evolved towards a state which, albeit not a true 
equilibrium, would persist as a near-equilibrium for times $>t_{H}$ in the
absence of any irregularities. However, such irregularities are always present,
and they could act to trigger a secular evolution, {\em e.g.,} driving the
system more nearly towards a true equilibrium.
\section{What are the effects of ongoing perturbations?}
Suppose that, after some more or less effective period of violent relaxation,
an elliptical has settled down {\em towards} (albeit not necessarily {\em to}) 
some realistic complex equilibrium or near-equilibrium. That configuration
could well be
distinctly nonaxisymmetric, at least in the outer regions, and, if so, might
be expected to exhibit variable axis ratios, becoming more nearly axisymmetric
near the center. However, the gravitational potential associated with such a
complex configuration is likely to involve significant measures of both 
regular and chaotic orbits. It is in fact well known to nonlinear dynamicists 
that less symmetric potentials tend to exhibit much larger measures of chaos, 
a point first stressed in the galactic dynamics community by 
\citeauthor{UaP} (\citeyear{UaP}).
Even comparatively simple potentials like the three-dimensional
logarithmic potentials, with 
$V={1\over 2}v_{0}^{2}\log (1 + x^{2}/a^{2}+ y^{2}/b^{2}+ z^{2}/c^{2})$,
admit chaotic orbits for certain choices of parameter values. Indeed, there
exist self-consistent axisymmetric equilibria which manifest chaotic meridional
motions, including the scale-free models considered by 
\citeauthor{nwe} (\citeyear{nwe}) 
and various spheroidal models considered by 
\citeauthor{hunt} (\citeyear{hunt}).

The important point, then, is that the phase space associated with a 
time-independent three-dimensional potential that admits significant measures 
of both regular and chaotic orbits tends to be very complex, being laced with 
cantori and/or an Arnold web. 
The existence of this complex structure implies that the approach
towards equilibrium can be comparatively inefficient. In particular, orbits
could well get trapped in localised phase space regions for very long times,
even though their motion is not blocked by a conservation law (like 
conservation of energy): In principle, an orbit may be able to access a large
phase space region, but it may have to leak through a narrow `bottleneck' to
get from one part of the accessible region to another, penetrating through
what the nonlinear dynamicist would call an {\em entropy barrier}.

This suggests the possibility that a system could evolve towards a state
which, albeit not a true equilibrium, could persist for times ${\gg}{\;}t_{H}$,
at least in the absence of significant perturbations. Indeed, this possibility
has been invoked by a number of authors 
({\em cf.} \citeauthor{paq} \citeyear{paq})
in the context of spiral galaxies, and corresponds to what 
\citeauthor{MaF} (\citeyear{MaF})
termed `quasi-equilibria' involving chaotic building blocks that
are only `partially mixed'. 

The important observation, however, is that such configurations can be 
surprisingly vulnerable to even very weak irregularities. This fact was 
apparently first recognised more than thirty years ago in the context of
simple maps ({\em cf.} \citeauthor{LaL1} \citeyear{LaL1});
and, over the past
several decades has proved to be very important in the context of plasma
physics and accelerator dynamics 
({\em cf.} \citeauthor{tenn} \citeyear{tenn}, \citeauthor{HaR} \citeyear{HaR}),
where one deals with beams of charged particles that are confined by
imperfect magnetic fields and in which discreteness effects, {\em i.e.,}
Rutherford scattering, can be surprisingly important.

But what can low amplitude perturbations actually do?  Even if the 
perturbations 
are too weak to significantly impact the values of the energy or any other 
collisionless invariants, they can induce systematic phase space flows 
on the (nearly) constant energy surfaces, {\em e.g.,} by allowing orbits 
originally trapped in localised phase space regions to become untrapped
({\em cf.} \citeauthor{Mah} \citeyear{Mah}, \citeauthor{KPS} \citeyear{KPS})
Moreover, under appropriate circumstances, perturbations can prove strong 
enough to induce nontrivial changes in the orbital energy and other 
collisionless invariants ({\em cf.} \citeauthor{kan01a} \citeyear{kan01a}).

Such perturbations act via a resonant coupling between the frequencies of
the perturbation and the frequencies of the orbits (\citeauthor{PaK}
\citeyear{PaK}). 
For regular orbits, the power is concentrated at a few special frequencies 
whereas, for chaotic orbits, power is typically broader band. In either case, 
however, when subjected to a perturbation characterised by its own set of 
natural frequencies, there is the possibility of resonance; and, to the extent 
that this resonance is strong, the perturbation will have a substantial effect

One might suppose that the response of the orbits to various influences
will depend sensitively on the precise form of the perturbation, so that 
believeable computations would require a knowledge of details that are 
difficult, if not impossible, to extract from observations. In point of fact,
however, the details seem comparatively unimportant. Because the perturbations
act via a resonant coupling, all that really seems to matter are the amplitude
of the perturbation, which determines how hard the orbits are being `hit,' and
the natural frequencies of the perturbation, which determine the degree to 
which a resonance is possible. 

The crucial point about this susceptibility to low amplitude irregularities is 
that changes in the orbital density and/or
collisionless invariants will in general yield changes in the bulk potential
and, as such, could trigger systematic evolutionary effects. It could be that
the system will react in such a fashion as to stabilise itself, but it might
equally well start to exhibit systematic changes in its bulk properties. 

The most detailed investigation of the effects of low amplitude perturbations
in the context of elliptical galaxies that has been effected hitherto 
(\citeauthor{SaK} \citeyear{SaK}, \citeauthor{KaS} \citeyear{KaS})
involved an investigation of flows in
the triaxial generalisations of the Dehnen potentials, with or without a 
central supermassive black hole. 
These potentials are not completely realistic since they assume strict 
ellipsoidal symmetry with fixed axis ratios, and ignore completely the
possibility of rotation. However, they do at least incorporate a high density
central concentration and, as might be expected physically, they include large
measures of both regular and chaotic orbits. 
Several different physical effects were considered:
\par\noindent${\bullet}$ {\em Discreteness effects}, {\em i.e.,} gravitational
Rutherford scattering between individual stars, were modeled as resulting in
dynamical friction and Gaussian {\em white noise}, {\em i.e.,} 
near-instantaneous impulses, in the spirit of 
\citeauthor{chan} (\citeyear{chan}). Even if
the relaxation time $t_{R}$ on which the energy of individual orbits changes
is long compared with $t_{H}$ which, as stressed by \citeauthor{MaF}
(\citeyear{MaF}), 
is not always so, such encounters can still be important by accelerating
diffusion on the nearly constant energy hypersurface 
({\em cf.} \citeauthor{HKM} \citeyear{HKM}, \citeauthor{KPS} \citeyear{KPS}).
\par\noindent${\bullet}$ The effects of {\em satellite galaxies and companion
objects} were modeled as {\em near-periodic driving}, characterised by a 
perturbation $V=V(\{{\omega}_{i}\}t)$, for some small number of frequencies 
$\{{\omega}_{i}\}$.
\par\noindent${\bullet}$ In {\em a dense cluster environment}, a galaxy will
be impacted by a large number of different neighbouring galaxies in a fashion
which is likely to be far from periodic. It thus seemed reasonable to 
model such an environment by allowing for {\em coloured noise}, this 
corresponding ({\em cf} \citeauthor{vanK} \citeyear{vanK}) to a
series of random impulses of finite duration. (Mathematically, coloured noise
can be viewed as a superposition of periodic disturbances with different
frequencies combined with random phases.)
\par\noindent${\bullet}$ In addition to companion objects, a galaxy can be
impacted by {\em tidal forces} associated with the bulk cluster potential 
which, depending on the form of the potential, could be {\em nearly periodic} 
or {\em largely random}.
\par\noindent${\bullet}$ {\em Coherent internal oscillations}, associated with
a small number of normal or pseudo-normal modes, were again modeled as inducing
a {\em near-periodic driving}. 
\par\noindent${\bullet}$ {\em Incoherent internal oscillations}, associated,
{\em e.g.,} with a large number of higher order modes, were again modeled as 
{\em coloured noise}, {\it i.e.,} near-random perturbations of finite duration.
(Making a sharp distinction between coherent and incoherent oscillations is 
admittedly somewhat {\em ad hoc}.)

Given the usual assumption that the noise under consideration is Gaussian, its
statistical properties are characterised completely by a knowledge of its 
first two moments. For both white and coloured noise, it was assumed that the
average force vanishes identically, so that ${\langle}F_{a}(t){\rangle}=0$ for
$a=x,y,z$. White noise, which involves instantaneous kicks, is characterised
by an autocorrelation function
\begin{equation}
{\langle}F_{a}(t_{1})F_{b}(t_{2}){\rangle}=2D\,{\delta}_{ab}
{\delta}_{D}(t_{1}-t_{2}).
\end{equation}
Coloured noise requires `fuzzing out' the Dirac delta. As a simple example,
this was done by sampling an Ornstein-Uhlenbeck process, for which
\begin{equation}
{\langle}F_{a}(t_{1})F_{b}(t_{2}){\rangle}=(D/t_{c}){\delta}_{ab}
\,\exp(-|t_{1}-t_{2}|/t_{c}).
\end{equation}
In each of these expressions, $D$ is the diffusion constant which, in the
white noise limit, enters into a Fokker-Planck description:
\begin{equation}
D{\;}{\equiv}{\;}\int_{0}^{\infty} 
\,d{\tau}\,{\langle}F_{a}(0)F_{a}({\tau}){\rangle}.
\end{equation}
The quantity $t_{c}$ is the autocorrelation time, which represents the
characteristic time scale on which the random forces change significantly.

The effects of low amplitude perturbations can be decomposed, at least
approximately, into (i) motions that involve little or no changes in the
energy and any other quantities which would be invariant for motion in a
fixed time-independent potential; and (ii) changes in the values of the
energy and any other collisionless invariants. 

Even if the perturbations are so weak that such collisionless invariants as
$E$ or $E_{J}$ are nearly conserved, they can significantly accelerate phase 
space
diffusion of chaotic orbits on the nearly constant energy surface. What this
entails is orbits leaking through topological partial obstructions associated 
with cantori and/or an Arnold web in a fashion strongly reminiscent of the
standard problem of {\em effusion}, whereby gas molecules can leak through a
tiny hole in a wall. If one considers an ensemble of `sticky' orbits 
({\em cf.} \citeauthor{con} \citeyear{con}) initially localised
in a given phase space region bounded by such obstructions, their escape is
well modelled as a Poisson process 
({\em cf.} \citeauthor{PaK} \citeyear{PaK}, \citeauthor{KPS} \citeyear{KPS}),
for which the number of orbits that 
remain trapped decreases exponentially, {\em i.e.,}
\begin{equation}
N(t)=N_{0}\exp(-{\Lambda}t).
\end{equation}
For a specified choice for the form of the perturbation, the value of 
${\Lambda}$ scales logarithmically in amplitude, so that, {\em e.g.,} for
the case of noise there is a logarithmic dependence on $D$. The effects
tend to be significant provided only that the characteristic time scale 
associated with the perturbation is not large compared with the natural
time scale associated with the orbits. For the case of coloured noise, 
the precise choice of autocorrelation time seems immaterial for $t_{c}<t_{D}$,
but the effects of the perturbation diminish significantly for $t_{c}{\;}
{\gg}{\;}t_{D}$. 

Under appropriate circumstances, the perturbations can also induce nontrivial
changes in such collisionless invariants as the energy. For the case of 
periodic driving, where the perturbations are characterised by a few special
frequencies, the detailed effects of the perturbation can be comparatively
complex ({\em cf.} \citeauthor{KAB} \citeyear{KAB}).
However, for the case of `random' disturbances, idealised as noise, 
the picture is quite simple, with the perturbations acting to trigger a 
{\em diffusion process}. For the case of white or near-white noise with 
$t_{c}{\;}{\ll}{\;}t_{D}$, the physics is identical to Brownian motion, as
proposed originally by \citeauthor{chan} (\citeyear{chan})
to model the effects of
gravitational Rutherford scattering between individual stars or by 
\citeauthor{SaS} (\citeyear{SaS})
to account for the scattering of disc stars off of giant
molecular clouds. For $t_{c}>t_{D}$ the effects once again begin to be 
suppressed. For the special case of coloured noise sampling the 
Ornstein-Uhlenbeck process, the {\em rms} change in energy is in general well
fit by a scaling relation of the form (\citeauthor{kan01a} \citeyear{kan01a})
\begin{equation}
{{\delta}E_{rms}\over |E|}{\;}{\sim}{\;} 
{\;}\left( { Dt\over |E| }  \right)^{1/2}\;\times\; \cases{
\;\;1 & for $t_{c}<t_{D}$\cr
 & \cr
\left( {t_{D}\over t_{c}} \right)  & for $t_{c}>t_{D}$.\cr}
\end{equation}
Overall, regular and chaotic orbits are affected in a nearly identical fashion,
except for very large values of $t_{c}$, where regular orbits prove to be
somewhat {\em more} susceptible than are chaotic orbits. 

The important point in all this is that realistic choices of $D$ and $t_{c}$
can actually have a significant effect. The diffusion constant scales as 
$D{\;}{\sim}{\;}F^{2}t_{c}$, where $F$ denotes the typical size of the random 
forces, so that its amplitude is readily estimated. 

On the nearly constant energy hypersurface, discreteness effects corresponding
to a relaxation time as long as $t_{R}{\;}{\sim}{\;}10^{6}-10^{7}t_{D}$ or
even more can have big effects within a time as short as $100\,t_{D}$. 
Similarly, even ignoring the effects of single very close encounters, random 
interactions with nearby galaxies in an environment where the separation 
between galaxies
is ten times the size of an individual galaxy, {\em i.e.,} 
$R_{sep}{\;}{\sim}{\;}10\,R_{gal}$, so that the autocorrelation time
$t_{c}{\;}{\sim}{\;}10\,t_{D}$, can have appreciable effects within 
$100\,t_{D}$
(\citeauthor{SaK} \citeyear{SaK}, \citeauthor{KaS} \citeyear{KaS}).

As regards changes in the energy and other collisionless invariants 
(\citeauthor{kan01a} \citeyear{kan01a}, \citeauthor{KaS} \citeyear{KaS}),
incoherent
internal oscillations at the $1\%$ level, far too small to be detected 
observationally, can trigger $10\%$ changes in the energy within $100\,t_{D}$;
and $10\%$ oscillations can trigger $30\%$ changes in a comparable time. 
Random interactions between neighbouring galaxies are less important, at least
directly, since an environment with $R_{sep}{\;}{\sim}{\;}6\,R_{gal}$ and 
$t_{c}{\;}{\sim}{\;}6\,t_{D}$ will only trigger $10\%$ changes within 
$100t_{D}$.
However, such interactions could still prove important 
({\em cf.} \citeauthor{VaW} \citeyear{VaW}) by triggering random incoherent 
oscillations that induce more substantial changes in the invariants.

\section{Discussion}

A major issue all too often unaddressed is the extent to which, as has been 
assumed here, real galaxies 
characterised by the highly irregular potential associated with a large number
of nearly point mass stars can in fact be approximated by a smooth 
three-dimensional potential. Should one, for example, really expect to see 
effects like `stickiness' or phase space diffusion, which have been studied 
primarily in the context of smooth two- and three-dimensional potentials, in
real many-body systems? Indeed, it
is clear that the continuum approximation misses at least some physical 
effects. It has, for example, become evident from both numerical computations
({\em cf.} \citeauthor{GHH} \citeyear{GHH}) and rigorous analytics 
(\citeauthor{Pogo} \citeyear{Pogo}) that, even for very large $N$, 
individual orbits in an $N$-body system typically have large positive Lyapunov 
exponents, even if the $N$-body system corresponds to an integrable density
distribution such as a spherical equilibrium. 

However, recent numerical work involving orbits and orbit ensembles evolved 
in fixed $N$-body realisations of continuous density distributions has 
shown that, in many respects, such `frozen-$N$' orbits are indistinguishable
from orbits evolved in the continuous density distribution in
the presence of friction and (nearly) white noise
(\citeauthor{KS} \citeyear{KS}, \citeauthor{SK} \citeyear{SK}).

Even though the Lyapunov exponents for frozen-$N$ orbits do not appear to
converge towards the values assumed by orbits in the potential associated
with the smooth density distribution, there is a precise sense in which,
as $N$ increases, frozen-$N$ trajectories remain `close to' smooth potential 
characteristics with the same initial conditions for progressively longer 
times. Viewed macroscopically, for both regular and chaotic initial conditions,
frozen-$N$ trajectories and smooth potential characteristics with the same 
initial condition typically exhibit a {\em linear} divergence: Their mean
separation ${\delta}r(t)$ is well fit by a growth law 
${\delta}r/R = A(t/t_{G})$, where $R$ represents the size of the configuration
space region accessible to the orbits and $A$ is a constant of order unity.
For regular orbits, $t_{G}/t_{D}{\;}{\propto}{\;}N^{1/2}$; for
chaotic orbits, $t_{G}/t_{D}{\;}{\propto}{\;}\ln N$. Moreover, for 
sufficiently large $N$ ensembles of frozen-$N$ orbits can exhibit `stickiness' 
qualitatively similar to what has been observed for orbits in
smooth potentials. And finally, both in terms of the statistical properties
of orbit ensembles {\em and} in terms of the pointwise properties of individual
orbits, discreteness effects associated with a frozen-$N$ system can be well
mimicked by white noise with a diffusion constant exhibiting (at least
approximately) the $N$-dependence predicted when these effects are modeled as
a sequence of incoherent binary encounters 
({\em cf.} \citeauthor{chan} \citeyear{chan}).

The extent to which the smooth potential approximation remains valid in 
the context of a self-consistent $N$-body evolution is much more difficult
to probe directly. However, recent work in the context of charged accelerator
beams, where particles interacting via a repulsive $1/r^{2}$ force are 
confined by an externally imposed potential, appears encouraging. In 
particular, fully self-consistent grid code simulations of intensed charged 
beams (which admittedly suppress the chaos associated with close encounters 
between individual particles) have been found to exhibit both regular and 
chaotic phase mixing qualitatively similar to what has been observed in smooth 
two- and three-dimensional potentials (\citeauthor{Kis} \citeyear{Kis}). 
Experiments to search for these effects in real particle beams are currently
in the planning stage (\citeauthor{Bohn} \citeyear{Bohn}) and analogous 
computations for self-gravitating systems 
are underway.

The principal message of this review is that it may be oversimplistic to
assume that elliptical galaxies should be viewed as collisionless equilibria.
Realistic equilibria, corresponding to nonaxisymmetric systems which contain a
high density central region, break strict ellipsoidal symmetry, and manifest
variable axis ratios, may not exist and, even if they do exist, may be very
difficult for real galaxies to find. Moreover, even if a galaxy seems `close
to' equilibrium observationally, it could well be comparatively `far from'
equilibrium from the standpoint of dynamics. Comparatively unsymmetric 
galaxies,
characterised by a complex phase space that admits a coexistence of significant
measures of both regular and chaotic orbits, can be surprisingly susceptible
to low amplitude perturbations of the form that act in the real world and, as
such, might be expected to exhibit systematic evolutionary effects over time
scales $<t_{D}$. What precisely these evolutionary effects could be is not
yet completely clear. However, there is good reason to believe that ongoing
observational programs, such as the Sloan Digital Sky Survey, will provide 
at least partial answers to this basic question.

\acknowledgements
I am pleased to acknowledge useful collaborations with Robert Abernathy,
Brendan Bradley, Barbara Eckstein, Salman Habib, Ilya Pogore-lov, Ioannis Sideris,
Christos Siopis, and, especially, Elaine Mahon, who first stimulated me to
think about the issues discussed in this paper. This research was supported in 
part by NSF AST-0070809 and by the Institute for Geophysics and Planetary 
Physics at Los Alamos National Laboratory.

\end{article}
\end{document}